\documentclass[10pt,journal]{IEEEtran}
\usepackage{amsmath,amsfonts}
\usepackage{algorithmic}
\usepackage[linesnumbered,ruled,lined]{algorithm2e}

\usepackage{array}
\usepackage[caption=false,font=normalsize,labelfont=sf,textfont=sf]{subfig}
\usepackage{textcomp}
\usepackage{stfloats}
\usepackage{url}
\usepackage{hyperref}
\usepackage{verbatim}
\usepackage{graphicx}
\usepackage{tabularx}
\usepackage{multirow}
\usepackage{amssymb}
\usepackage{booktabs}
\usepackage{cite}
\usepackage{color}
\usepackage[utf8]{inputenc}
\usepackage{tabularx} 
\usepackage{enumitem}
\hypersetup{
    hidelinks
}

\begin{document}

\title{Boundary-Aware Multi-Behavior Dynamic Graph Transformer for Sequential Recommendation}

\author{
  Jingsong Su, Xuetao Ma, Mingming Li*, Qiannan Zhu*, Yu Guo \IEEEcompsocitemizethanks{\IEEEcompsocthanksitem Jingsong Su, Xuetao Ma, Qiannan Zhu and Yu Guo are with School of Artificial Intelligence, Beijing Normal
University, Mingming Li is with Department of Search and Recommendation at JD.com, Beijing. \protect\\
E-mail: sujingsong@mail.bnu.edu.cn, zhuqiannan@bnu.edu.cn
\IEEEcompsocthanksitem Mingming Li and Qiannan Zhu are the corresponding author.}%
\thanks{Manuscript received April 19, 2021; revised August 16, 2021.}
}


\markboth{Journal of \LaTeX\ Class Files,~Vol.~14, No.~8, August~2021}%
{Shell \MakeLowercase{\textit{et al.}}: A Sample Article Using IEEEtran.cls for IEEE Journals}


\maketitle

\begin{abstract}

In the landscape of contemporary recommender systems, user-item interactions are inherently dynamic and sequential, often characterized by various behaviors. Prior research has explored the modeling of user preferences through sequential interactions and the user-item interaction graph, utilizing advanced techniques such as graph neural networks and transformer-based architectures. However, these methods typically fall short in simultaneously accounting for the dynamic nature of graph topologies and the sequential pattern of interactions in user preference models. Moreover, they often fail to adequately capture the multiple user behavior boundaries during model optimization. To tackle these challenges, we introduce a boundary-aware Multi-Behavioral Dynamic Graph Transformer (MB-DGT) model that dynamically refines the graph structure to reflect the evolving patterns of user behaviors and interactions. Our model involves a transformer-based dynamic graph aggregator for user preference modeling, which assimilates the changing graph structure and the sequence of user behaviors. This integration yields a more comprehensive and dynamic representation of user preferences. For model optimization, we implement a user-specific multi-behavior loss function that delineates the interest boundaries among different behaviors, thereby enriching the personalized learning of user preferences. Comprehensive experiments across three datasets indicate that our model consistently delivers remarkable recommendation performance.

\end{abstract}


\section{Introduction}
\IEEEPARstart{W}{ith} the exponential growth in online information, 
recommender systems have evolved into indispensable instruments to navigate the deluge of information across various services, including e-commerce\cite{rec-for-eco,rec_for_eco2,example_for_eco,li2023adaptive,li2024generativeretrievalpreferenceoptimization,kuai-etal-2024-breaking,liqueryindex,li-etal-2024-shot,lirerank,liMM}, video streaming\cite{rec-for-video, rec_for_video2}, and online retailing platforms~\cite{DIPN, GMV1, GMV2}. These systems are designed to predict items that users may find appealing based on their individual preferences, which can be inferred from a range of data such as the behavior of similar users, historical interactions, and direct feedback like ratings or reviews. By analyzing this information, recommender systems could offer tailored suggestions, improving the user experience by connecting them with pertinent content or products. 


In recent years, sequence-based methods \cite{GRU4Rec,Caser,SASRec,Bert4rec} have risen to prominence and achieved promising performance in personalized recommendation systems. These methods leverage the sequential interactions of users, capturing the temporal order and context of user behavior to better predict future preferences. By modeling the sequence of actions, such as clicks, purchases, or views, these approaches can provide more accurate and timely recommendations, significantly enhancing the overall user experience.  
However, these widely adopted recommendation models tend to simplify the complexity of user interactions by focusing on just one type of behavior, thus overlooking the rich diversity of ways users interact with items.

In real-world scenarios, multiple interaction behaviors, such as adding items to favorites and to the cart, provide diverse feedback and reflect user preferences more comprehensively. Moreover, considering multiple-behavior interaction sequences can alleviate the sparsity problem of the target purchase behavior for more accurate recommendations. To fully utilize multi-behavioral information, a large amount of work on multi-behavior sequential recommendation has been proposed, which can be roughly classified into two branches: 
i) A series of studies\cite{DIPN, example_first_class, MGNN-SPerd, DMT} decompose the multi-behavior interaction sequences into several single-behavior interaction sequences for later joint modeling, 
isolating co-influence across different behavior-specific sub-sequences. For instance, DIPN\cite{DIPN} models buying and browsing separately using a bi-directional RNN and attention mechanism. DMT\cite{DMT} uses multiple transformers for different behaviors and combines them with a multi-gate mix -of- expert. 
ii) Another branch of work\cite{ example_sec_class2, MBHT, MB-STR} independently models item and behavioral sequences, considering behavioral knowledge as a supplementary aspect of the representation. Recent work in this branch of the approach injects behavioral interaction context into item transitions, for example, MBHT\cite{MBHT} utilizes a multi-scale transformer and hypergraph for behavior-aware item transitions. MB-STR\cite{MB-STR} models heterogeneous item-level multi-behavioral dependencies with a transformer layer.

\begin{figure}
    \centering
    \includegraphics[width=1\linewidth]{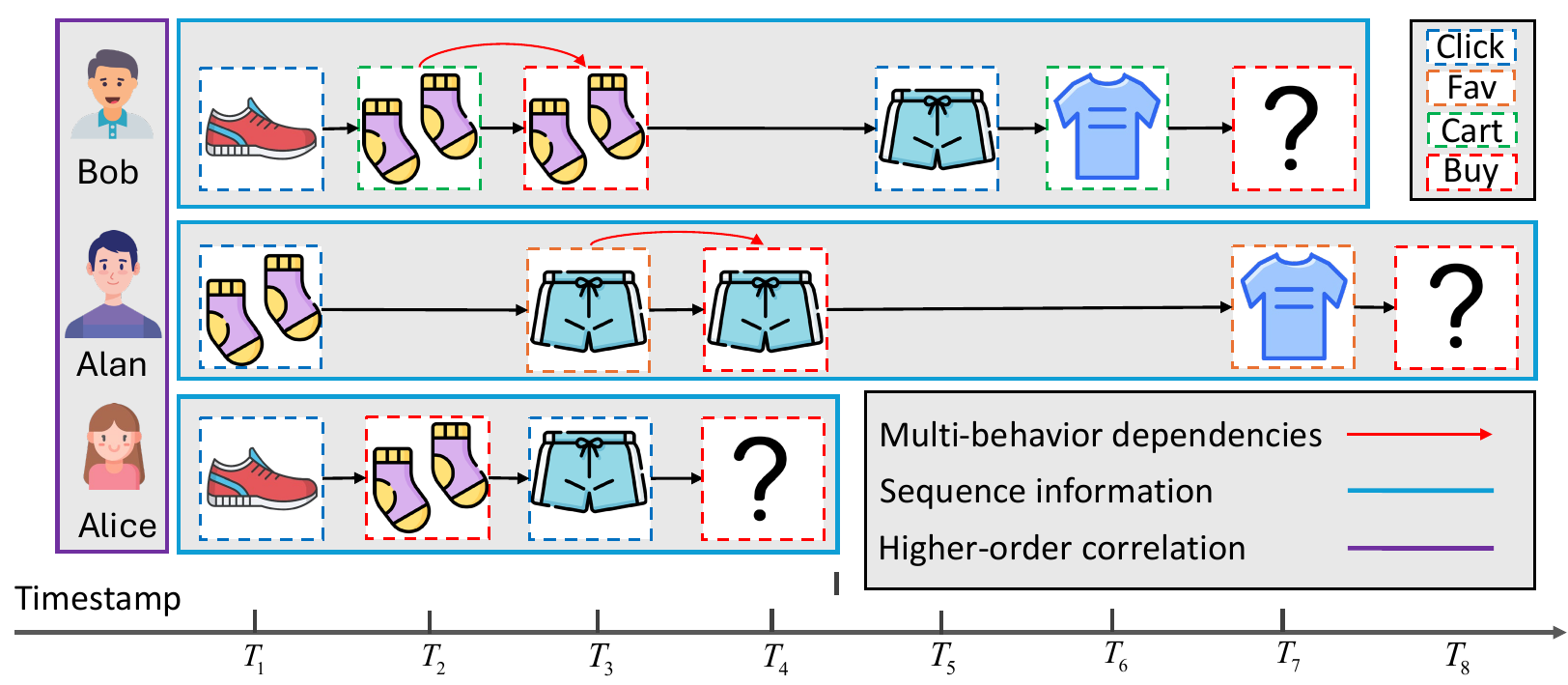}\\
    \caption{This figure illustrates multi-behavior sequential recommendation, depicting interactions across timestamps for three users. }\label{fig:fig1}
  \end{figure}

Despite their effectiveness, firstly, a number of these methods primarily capture users' interests within their own sequences, neglecting the higher-order correlation between different user/item sequences\cite{DGNN_for_sequential}.
As Fig~\ref{fig:fig1}. illustrates, Alice's interaction items before moment $T_4$ have also been interacted with by Bob and Alan, suggesting potential similarities in their preferences at different times. 
Similarly, higher-order connectivity signals such as category attributes exist between items, e.g., items of the same brand, category, or often purchased together. Ignoring higher-order correlations between users and items often leads to poor model performance in cold-start situations with sparse data, necessitating longer sequences to compensate. Secondly, although existing methods effectively model multi-behavioral dependencies in interaction sequences, they struggle to capture dynamic changes in user interests.  Previous studies have utilized positional encoding, with few examining the direct impact of time intervals on behavior-aware collaborations. Intuitively, a shorter interval suggests a higher degree of collaboration.
Neglecting the impact of time intervals makes it difficult to timely capture dynamic changes in user interests and may introduce noise into the recommendations\cite{noise}. Thirdly, users with varying interests have distinct interest boundaries\cite{boundary}. Previous model optimization methods using point-wise or pair-wise\cite{pair_point} loss functions lack personalization, making it difficult to establish a learnable personalized decision boundary for user behaviors to identify positive samples during the ranking stage.

\begin{figure}
    \centering
    \includegraphics[width=1\linewidth]{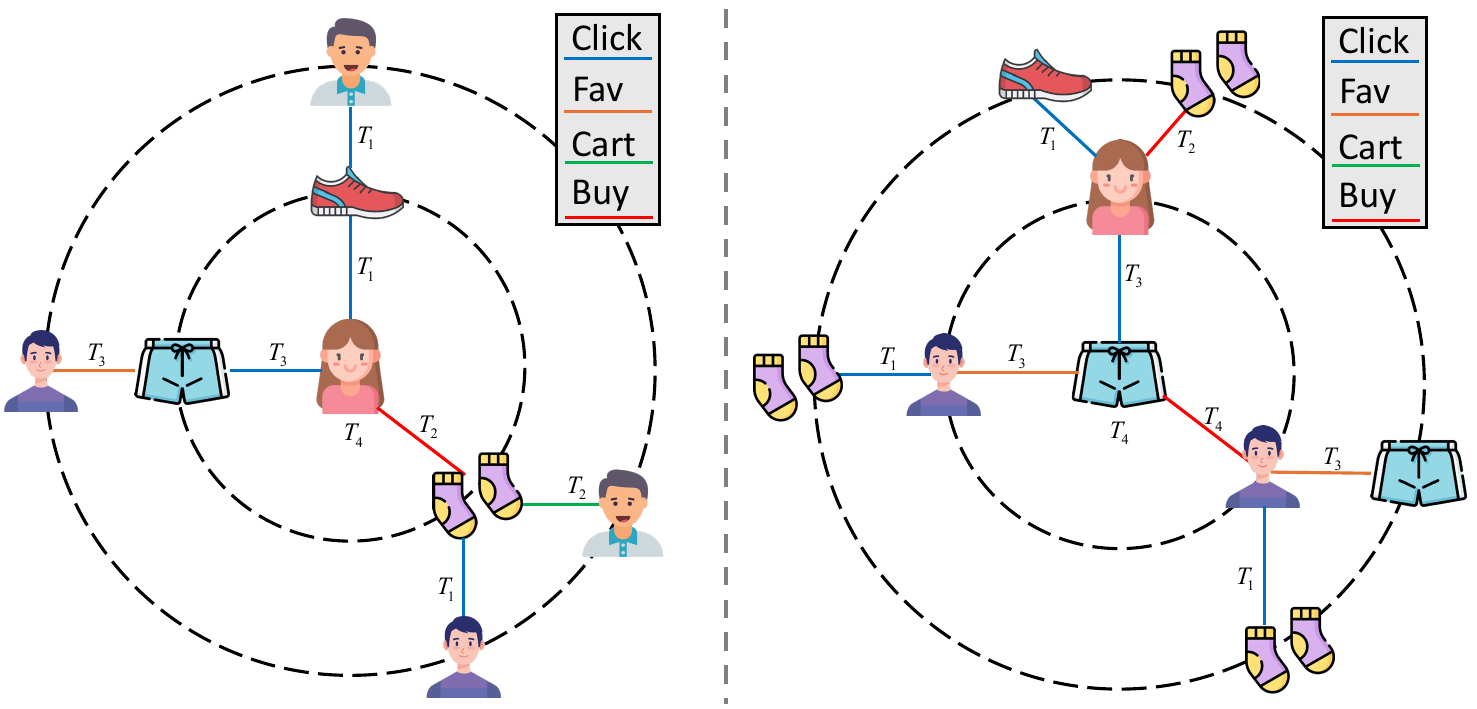}\\
    \caption{Two examples of 2-hop multi-behavior temporal interaction graphs with user and item as target nodes respectively at moment $T_4$}\label{fig:fig2}
  \end{figure}


To address the aforementioned issues, we propose a boundary-aware Multi-Behavior Dynamic Graph Transformer (MB-DGT) model for addressing the challenge of multi-behavior sequential recommendation. 
Our approach begins by deploying a transformer architecture specifically designed to unravel the complex dependencies that exist within sequences of user behaviors. This is coupled with an $L$-hop graph structure that adeptly captures the intricate higher-order interactions among user sequences. This dual strategy is pivotal in dynamically shaping user interest profiles through both behavior-sensitive encoding and time-aware graph-based aggregation. As Figure~\ref{fig:fig2} illustrates, we construct two types of multi-behavior temporal interaction graphs: one that traces the user-item-user trajectory and another that follows the item-user-item trajectory, placing users and items at the core of these graphs. We initiate the process by encoding neighboring nodes via a behavior-aware transformer to map out the multi-behavioral dynamics, which is then followed by a temporal-aware aggregation to craft the final user/item embedding.
Furthermore, to overcome the last limitation, we introduce a learnable explicit score boundary for every user's behavior in the loss function. This approach penalizes positive samples with scores below the boundary and negative samples with scores above the boundary. By using the hyperparameter $\alpha$ to control the score distribution, we can explore differences in interest boundaries across various user behaviors, achieving a more accurate and personalized recommendation. We perform experiments across three available datasets to validate the performance of our MB-DGT framework. Additionally, ablation studies highlight the contribution of each component within our framework.

Our principal contributions are outlined as follows:

\begin{itemize}
    \item We propose a new framework MB-DGT for multi-behavior sequential recommendation, uncovering the dynamics of user interests and multi-behavioral dependencies in user-item interactions.
    \item We employ a transformer-based multi-behavioral correlation model for behavior-aware sequence modeling, temporal graph aggregation to capture user interests at intermediate moments, and an $L$-hop graph structure to capture higher-order correlations within users and items.
    \item In addition, we introduce a learnable explicit score in the loss function to represent the user's interest boundaries across different behaviors, further enhancing recommendation effectiveness by adjusting the boundary distribution.
\end{itemize}

\section{Related Work}
In this section, we examine recent studies on sequential recommendation, multi-behavior recommendation, and multi-behavior sequential recommendation.
\subsection{Sequential Recommendation}
Sequential recommendation methods predict the next items users will interact with by leveraging the sequential dependencies found in their historical interactions.
Early work\cite{markov, marrkov2} often rely on Markov chain assumptions
to model sequential dependencies. 
As deep learning has advanced rapidly, there has been a substantial focus on developing sequential recommendation systems enhanced by neural networks to capture the intricate dependencies among items from multiple perspectives.
For example, GRU4Rec\cite{GRU4Rec} employs GRUs to model temporal information in user sessions. Caser\cite{Caser} uses temporal convolutional neural networks to capture the user's both long-term and short-term preferences for recommendation. With the rising popularity of the Transformer architecture, SASRec\cite{SASRec} and BERT4Rec\cite{Bert4rec} employ self-attention mechanisms to model user behavioral sequences for recommendations. In addition, some methods use GNNs to perform message passing between neighboring items, capturing sequential signals. By propagating embeddings across the user-item graph, NGCF\cite{NGCF} leverages the user-item graph structure to effectively model high-order correlations within the graph. Recently, contrastive learning-based models\cite{example_for_contrastive,example_for_contrastive2,example_for_contrastive3,example_for_contrastive4} have been developed to identify significant user patterns by utilizing self-supervised signals. However, most of these approaches primarily address only one type of behavior and fail to manage diverse user-item behavior relationships.


\subsection{Multi-Behavior Recommendation}
Multi-behavior recommendation methods improve retrieval performance for the target behavior by leveraging different interaction types. Early studies\cite{juzhenfenjie1, juzhenfenjie2, juzhenfenjie3} on multi-behavior recommendation primarily use matrix factorization methods. They employ shared user or item representations to simultaneously decompose multiple behavior-specific interaction matrices.  Later, some approaches\cite{NMTR, MATN, example_for_rnn,example_for_mbr} use neural networks or transformers to model behavioral dependencies in multi-behavior interaction sequences. For example, NMTR\cite{NMTR} proposes a new neural network model that uses a multi-task learning paradigm to capture associations between different behaviors in a cascading manner.
MATN\cite{MATN} uses a transformer to represent interactions across multiple behaviors and introduces a memory-enhanced attention network to map contextual signals of different behavior types into distinct latent embedding spaces. Furthermore, some approaches\cite{MB-GCN, MB-GMN, KHGT, example_for_GNN, example_for_GNN2} leverage graph neural networks (GNNs) to model higher-order multi-behavioral interactions through message passing on graphs. For example, MB-GCN\cite{MB-GCN} refines user/item representations by passing behavior-aware messages over the user-item interaction graph using a multi-behavioral GCN network. Besides, MB-GMN\cite{MB-GMN} 
designs a meta-graph framework to identify unique multi-behavior signals and represents the various dependencies among different behaviors.
KHGT\cite{KHGT} integrates GNN and Transformer to capture higher-order connectivity and multi-behavioral dependencies in user-item interaction graphs. Recently, some methods\cite{CML, VAE++, multi-interest1,li2020symmetric} have also achieved promising results using contrastive learning, variational autoencoder or multi-interest learning. CML\cite{CML} encodes the heterogeneity of different user behaviors using a contrastive meta-network. VAE++\cite{VAE++} proposes a novel variational autoencoder-based recommendation model utilizing heterogeneous feedback to improve recommendation performance. Although the heterogeneity of user-item interactions is considered, there is an inherent order in the user's implicit feedback that the above approaches do not fully utilize, neglecting the sequential information.

\subsection{Multi-Behavior Sequential Recommendation}
Multi-behavior sequential recommendation methods incorporate users' heterogeneous behavior data and the sequential information within and across behaviors, thereby more accurately reflecting real-world recommendation scenarios.
Existing works on multi-behavior sequential recommendation can be classified into two main categories. 
The first type of approaches\cite{DIPN, example_first_class, MGNN-SPerd, DMT, example_first_class2} divides a multi-behavior interaction sequence into behavior-specific subsequences, modeling both the sequential information within each subsequence and the dependencies between them. For instance, DIPN\cite{DIPN} proposes a hierarchical attention mechanism to learn correlations between various behaviors. DMT\cite{DMT} uses a Transformer with unshared parameters to simultaneously model users' various behavior sequences. MGNN-SPred\cite{MGNN-SPerd} explores global item-to-item correlations by constructing a multi-faceted item graph and treating two sequences of user behaviors as two sub-graphs within it. Although these approaches independently model the semantics of different behavior types by modeling sequences of behavior-specific items, they overlook multi-behavior dependencies within the sequential context. The second type of approaches \cite{ example_sec_class2, MBHT, MB-STR} models item sequences and corresponding behavioral sequences separately, which either utilizes the sequence of behavior to enrich the model input or 
explicitly represents the behavior sequences.
For example, BINN\cite{BINN} constructs a context-aware LSTM that incorporates behavioral information as input to memorize the multi-behavioral information in sequences. Recent approaches\cite{MBHT, MB-STR} suggest incorporating behavioral information into item representations. MBHT\cite{MBHT} designs a hypergraph-enhanced multi-scale transformer to capture the sequential patterns and capture global dependencies across various behaviors. MB-STR\cite{MB-STR} designs a multi-behavior transformer to model heterogeneous item-level dependencies and simulates multi-behavioral sequence relationships using a novel positional encoding function. This type of approach\cite{  example_sec_class2, MBHT, MB-STR} enables models to differentiate between various behavior types while preserving the integrity of the entire interaction sequence. In this study, our framework continues to advance the second approach to multi-behavior sequential recommendation.

\section{Preliminaries}
This section introduces the basic concept of the multi-behavior sequential recommendation and multi-behavior temporal interaction graph, followed by a detailed problem formulations. The variables used in this paper are summarised in Table~\ref{tab:notations}.
\subsection{Task Definition}
In our multi-behavioral sequential recommendation scenarios, \( \mathcal{U} = \{u_1, u_2, \dots, u_{|\mathcal{U}|}\} \) denotes user set and \( \mathcal{V} = \{v_1, v_2, \dots, v_{|\mathcal{V}|}\} \) denotes the item set, where the number of users is represented by \(\lvert \mathcal{U} \rvert\), the number of items is represented by \(\lvert \mathcal{V} \rvert\). Given a user $u_i \in \mathcal{U}$, the sequential interactions with items are multi-behavioral and temporally ordered. So we define \( \mathcal{B} = \{b_1, b_2, \dots, b_{|\mathcal{B}|}\} \) and \( \mathcal{T}_{u_i} = \{t_1, t_2, \dots, t_n\} \) denote the set of behaviors and the timestamps at which user $u_i$ interacts with the item, where the number of behavior types is represented by \(\lvert \mathcal{B} \rvert\) and $n$ represents the sequence length. Then the multi-behavioral interaction sequence $S_{u_i}$ can be represented as $S_{u_i}=\left[\left(v_{1}, b_{1}, t_{1}\right), \ldots,\left(v_{k}, b_{k}, t_{k}\right), \ldots,\left(v_{n}, b_{n}, t_{n}\right)\right]$, where $v_{k}\in \mathcal{V}$, $b_{k}\in \mathcal{B}$ and $t_{k}\in \mathcal{T}_{u_i}$ respectively denote the item, behavior, and timestamp of the $k$-th interaction of user $u_i$. We pad the sequence with zeros if the original sequence length fall short of $n$. Furthermore, the target behavior is defined as purchasing, while the remaining behaviors serve as auxiliary behaviors.


With the above definitions, the task of multi-behavior sequential recommendation can be characterized as:
\textit{Input}: the multi-behavior interaction sequence $S_{u_i}$ $\left(u_i \in U\right)$. \textit{Output}: the learned probability that user $u_i$ engages with item $v_{n+1}$ in the target behavior at the $t_{n+1}$ timestamp.

\subsection{Multi-Behavior Temporal Interaction Graph}
In our work, the user multi-behavioral interaction sequences can be defined as a cascading multi-behavioral temporal interaction graph $\mathcal{G}=(\mathcal{N}, \mathcal{E}, \mathcal{T})$, where $\mathcal{N}$ denotes a user/item node, $\mathcal{E}$ and $\mathcal{T}$ denote the attributes of interaction edges, indicating different types of behaviors and the timestamps corresponding to the interactions, respectively.
Our research primarily focuses on the concept of continuous-time dynamic graphs\cite{TGN, TGAT}. The node and graph topology features continually change over time. By temporally modeling multi-behavioral interaction sub-graphs at intermediate moments within a sequence of user interactions, we can divide the user multi-behavioral interaction graph $\mathcal{G}$ into multiple sub-graphs ${\mathcal{G}_{t_1}, \ldots, \mathcal{G}_{t_k}, \ldots, \mathcal{G}_{t_n}}$ at different timestamps, where $\mathcal{G}_{t_k}$ refers to the sub-graph modeled in the interaction sequence prior to the timestamp $t_k$. Fig~\ref{fig:fig2}. illustrates our constructed graph. The notation table is in TABLE~\ref{tab:notations}.

\begin{table}[t]\centering
    \caption{Notation Summary used in this paper.} \label{tab:notations}
    \begin{tabular}{ l | l  }
      \toprule
      \textbf{Symbol}&\textbf{Description}\\
      \midrule
      $\mathcal{U}$ & the user set\\
      $\mathcal{V}$ & the item set\\
      $\mathcal{B}$ & the behavior set\\
      $\mathcal{T}_{u_i}$ & the set of user-item interaction timestamps\\
      $S_{u_i}$ & the multi-behavior interaction sequence\\
      $\left(v_{k}, t_{k}, b_{k}\right)$ & the interaction instance in the k-th sub-sequence for $u_i$\\
      $n$ & the sequence length\\
      $d$ & the model dimension\\
      $t$ & the timestamp of the interaction\\
      $k$ & the k-th interaction\\
      $h$ & the number of heads\\
      $E(v)$ & the initial embedding of neighbor sequences\\
      ${E^l}(v)$ & the behavior correlation-aware node embeddings\\
      $z(t)$ & the time- and behavior-aware node embeddings \\
      ${h^{(L+1)}}(t)$ & the aggregated node embeddings\\
      $b_{u,k}(t)$ & the user's k-th interaction behavior boundary at t\\
      $\mathcal{G}$ & the muli-behavior temproal interaction graph\\
      $\mathcal{N}$ & the user or item node\\
      $\mathcal{E}$ & the behavioral types of interactions\\
      $\mathcal{T}$ & the timestamps of the corresponding interactions\\
      $\mathcal{G}_{t_k}$ & the sub-graph prior to the timestamp $t_k$\\
    \bottomrule
  \end{tabular}
  \end{table}

\section{The Proposed Method}
\begin{figure*}[htbp]
  \centering
  \includegraphics[width=0.952\linewidth]{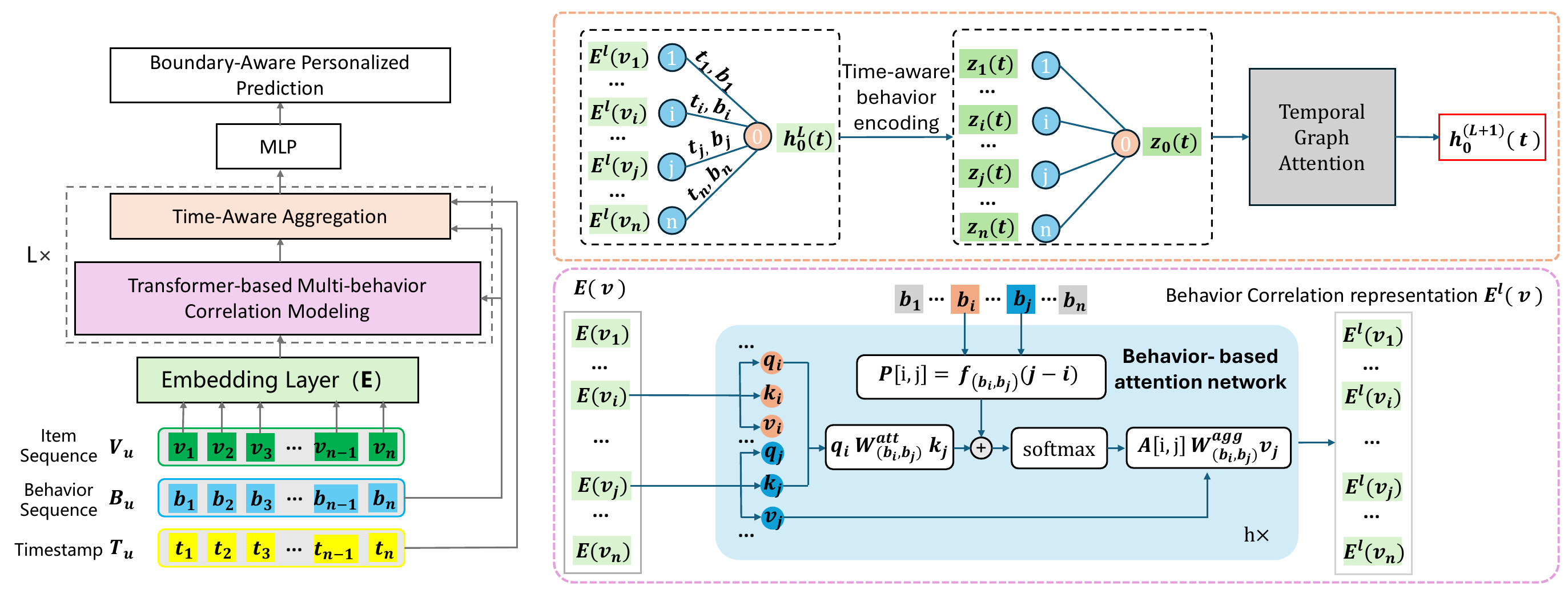}
  \caption{The model architecture of MB-DGT with users as target nodes. The overall structure is depicted in the left panel, while the right panel provides a detailed view of the model's two main modules. The lower right corner illustrates the serialized representation encoding process for behavioral association perception, and the upper right corner demonstrates the time-aware aggregation process for one-hop neighbors.}\label{fig:method}
\end{figure*}

In this section, we first present an overall description of our model. 
Next, we offer an in-depth explanation of each component. Finally, we discuss the complete algorithm and perform a complexity analysis.


\subsection{Overview}
We now introduce our framework MB-DGT.  As Fig~\ref{fig:method} illustrates, there are three main components: (i) the Transformer-based Multi-behavior Correlation Modeling(TMCM) module; (ii) the Time-Aware Aggregation(TAA) module; and (iii) the Boundary-Aware Personalized Prediction(BAPP) module. Our method processes $L$-hop user-item multi-behavior temporal interaction graphs by stacking TMCM module and TAA module $L$ times. We first embed the sequence of neighbors of the target node on the graph through an embedding matrix $ E \in R^{(|\mathcal{V}| + |\mathcal{U}|) \times d}$. Then we generate behavioral correlation-aware neighbor node representations within sequences through the TMCM module. Subsequently, the neighbor node representations and the behavioral and temporal attributes of the interacting edges in the graph are aggregated through the TAA module to generate the target node representations. Finally, the behavioral boundaries are modeled by the BAPP component for loss optimization and predict items that are probably to interact in the subsequent moment. Next, we will elaborate on these three components.

\subsection{Transformer-based Multi-behavior Correlation Modeling}

We encode the sequence of neighbor interactions for the target node up to timestamp $t_k$ as a multi-behavior-aware sequence embedding. Specifically, we capture attentional dependencies between behaviors using multi-behavioral multi-head self-attention\cite{transforrmer} and inject sequential signals into the behavior sequences through multi-behavioral relative positional encoding. We describe these two components in detail below:


\subsubsection{\textit{Multi-Behavioral Multi-head Self-Attention}} Consider the situation in Fig~\ref{fig:method} as an example, where the sequence of items \( V_u = \{v_1, v_2, \dots, v_n\} \) from user interactions first passes through the embedding layer $ E \in R^{(|\mathcal{V}| + |\mathcal{U}|) \times d}$ to obtain the $d$-dimensional initial representation \( E(v) = \{E(v_1), E(v_2), \dots, E(v_n)\} \).

Then, we use $E(v)$ as input and introduce one-hot encoding of behavioral sequences for behavior-specific linear mapping \(R^{n \times d} \rightarrow R^{n \times \frac{d}{h}}\) to generate $q$, $k$ and $v$. This approach allows the linear projection to adjust weights according to behavioral categories to better capture the differences and associations between behaviors, which can be represented as:
\begin{equation}\label{eq:qkv]}
  {\{q, k, v\} = \{f_{q}, f_{k}, f_{v}\}(E(v), b)}
\end{equation}

Next, in calculating the raw attention scores, the matrix $W^{att}_{(b_i, b_j)} \in R^{\frac{d}{h} \times \frac{d}{h}} $ introduces behavior-specific attention weights, allowing different behavioral categories to have distinct attention scores. For behavioral sequences, we add relative position encoding to incorporate behavioral sequence information, further details are provided in the next subsection. The finalized attention weight can be denoted as:
\begin{equation}\label{eq:attention weights}
  A[i, j] = \text{Softmax} (\frac{q[i] W^{att}_{(b_i, b_j)} k[j]}{\sqrt{d}} + P[i,j])
\end{equation}

Furthermore, in calculating the final attention output, behavioral one-hot encoding and behavioral weighting matrices are introduced, and a matrix $W^{agg}_{(b_i, b_j)} \in R^{\frac{d}{h} \times \frac{d}{h}} $ with different parameters is constructed for each behavioral pair to obtain the final attention output. The formula can be expressed as:
\begin{equation}\label{eq:final attention output}
  S[i] = \sum_j A[i,j] W^{agg}_{(b_i, b_j)} v_j
\end{equation}
where $S[i]$ represents the attention score for each position in a sequence of length $n$. Given that we use a $h$-head self-attention mechanism, we concatenate the outputs from all $h$ heads to obtain the representation ${S^l}[i]$, as expressed by the following equation:
\begin{equation}\label{eq:s^l[i]]}
  {S^l}[i] = \text{Concat}( {S_1}[i], \ldots, {S_h}[i])
\end{equation}

Finally, we design $|\mathcal{B}|$ MLPs with different parameters for various types of behavioral interactions. We derive the final behavioral correlation representations through the corresponding behavior-specific MLPs. The equation can be expressed as:
\begin{equation}\label{eq:final representations]}
  {E^l}(v_i) = \text{MLPs}( {S^l}[i], b)
\end{equation}
where $E^l(v) = \{{E^l}(v_1), {E^l}(v_2), \ldots, {E^l}(v_n)\}$ is the sequence representation of neighbor nodes with behavioral correlations. Residual connections\cite{residual} and layer normalizations\cite{layernorm} are also utilized to get the final representation of neighbor nodes with behavioral correlations:
\begin{equation}\label{eq:Ln and residual}
  \begin{aligned}
    {S^l}[i] &= \text{LayerNorm}(E(v) + {S^l}(i)), \\
    {E^l}(v) &= \text{LayerNorm}({S^l}(i) + {E^l}(v))
  \end{aligned}
\end{equation}

\subsubsection{\textit{Multi-Behavior Relative Positional Encoding}} User interactions often follow sequential patterns; for example, clicking behaviors usually represent short-term interests, affecting user interactions only for a relatively short period, while purchasing behaviors indicate long-term interests, influencing current interactions even if the purchase occurred a long time ago. To capture these sequential patterns, we introduce relative position encoding\cite{T5} for multi-behavior sequences, formally represented as:
\begin{equation}\label{eq:P[i,j]}
  {P[i, j] = f_{(b_{i}, b_{j})}(j - i)}
\end{equation}

Specifically, $P[i,j]$ and the original attention score share the same dimension. $P[i,j]$ quantifies the extent to which position $i$ should focus on position $j$, based on their relative position $(j-i)$ and their behavior types $(b_i,b_j)$. By encoding the behaviors at positions $i$ and $j$ pairwise in the sequence, we can capture the sequential patterns between the behaviors.

To choose the position encoding method, we consider that mapping relative positions directly to real numbers can lead to uneven learning, especially for long relative positions. For example, in a sequence of length 30, there are 29 pairs of positions with a relative distance of 1, while there is only 1 pair with a relative distance of 29. Therefore, we refer to the T5\cite{T5} model for a finer segmentation of relative positions. Specifically, each position is assigned an independent position when the relative distance is short; as the relative distance increases, multiple positions share a position until the relative distance reaches a threshold and is truncated. This approach effectively equalizes the number of short-distance position pairs and allows for a more balanced modeling of the long length behavioral sequences. The specific segmentation processing formula is expressed as:
\begin{equation}\label{eq:f(i,j)}
  {f(i,j) = \begin{cases} 
    \text{if } 0 \leq i-j < \frac{n_b}{4}:  i-j \\
    \text{if } -\frac{n_b}{4} \leq i-j < 0:  i-j+\frac{n_b}{2} \\
    \text{if } i-j \geq \frac{n_b}{4}: & \\ \min\left( \frac{n_b}{4}+\lfloor{\frac{\log 2(i-j) - \log \frac{n_b}{2}}{\log 2n -\log \frac{n_b}{2}}}\rfloor \cdot (\frac{n_b}{4}), \frac{n_b}{2}-1\right) \\
    \text{if } i-j \leq -\frac{n_b}{4}: & \\ \min\left( \frac{3n_b}{4}+\lfloor{\frac{\log 2(i-j) - \log \frac{n_b}{2}}{\log 2n -\log \frac{n_b}{2}}}\rfloor \cdot (\frac{n_b}{4}), n_b-1\right) 
    \end{cases}}
\end{equation}
where $i$ and $j$ are the indexes of a pair of behaviors in the behavior sequence, $n_b$ is the number of positional encodings after the segmentation process, and $n$ is the maximum relative distance, i.e., the length of the sequence.

\subsection{Time-Aware Aggregation}
In this section, we will complete the time-aware aggregation of target nodes using the neighbor sequence representations obtained earlier. First, we use time-aware behavior encoding to assign temporal and behavioral information to the neighbor sequence node representations. Then, the time-spanning and behavior-aware neighbor sequence node representations are aggregated to the target node through the temporal graph attention mechanism. These two components are described in detail below:

\subsubsection{\textit{Time-aware behavior encoding}} We start by defining a continuous function $\Phi : T \rightarrow \mathcal{R}^{d_T}$ that maps the temporal range into a $d$-dimensional vector space representation, serving as an alternative to traditional positional encoding. Let us consider two specific time points, $t_1$ and $t_2$, and examine the inner product of their encoded functions, $\langle \Phi(t_1), \Phi(t_2) \rangle$.
Our primary focus is on learning patterns associated with the timespan $|t_2 - t_1|$. Ideally, this relationship should be captured by $\langle \Phi(t_1), \Phi(t_2) \rangle$  to ensure compatibility with self-attention mechanisms.

This study\cite{TGAT} introduces a continuous, translation-invariant kernel $K(t_1, t_2)$ and, with reference to Bochner's theory and Monte Carlo integral\cite{Bochner}, deduces that there exists a continuous time mapping function $\Phi(t)$:
\begin{equation}\label{eq:t}
  {\Phi(t) = \sqrt{\frac{1}{d}} \left[ \cos(\omega_1 t), \sin(\omega_1 t), \ldots, \cos(\omega_d t), \sin(\omega_d t) \right]}
\end{equation}
satisfying the condition:
\begin{equation}\label{eq:K}
  {\Phi(t_1)^T\Phi(t_2) \approx \varphi(t_1 - t_2) = K(t_1, t_2)}
\end{equation}
which aligns well with our starting point and can be used as a time-aware encoding function to incorporate time-span information into the attention mechanism.

After embedding temporal features for nodes, it is also necessary to embed behavioral features on the edges of the interaction graph. For this purpose, we embed the different kinds of interaction behaviors as a feature vector $x_{i,j}(t) \in \mathcal{R}^{d_B}$, where $x_{i,j}(t)$ denotes the embedding of the type of behavior of node $i$ interacting with node $j$ at time $t$. By combining the original node representations with these behavioral and temporal representations, our time-aware behavioral encoding as shown in Fig~\ref{fig:method}. can be expressed as:
\begin{equation}\label{eq:Z1(t)}
  {z_i(t) = \left[ E^{l}(v_i) \| x_{0,i}(t_i) \| \Phi(t - t_i) \right]  }
\end{equation}

Finally, through time-aware behavior encoding, we successfully incorporate temporal and behavioral information into the target and neighbor node representations before aggregation.

\subsubsection{\textit{Temporal Graph Attention Layer}} First, we introduce the preparatory work before temporal graph aggregation. To aggregate the neighbor node representations and obtain the target node's representation at time $t$, we use node 0 in Fig~\ref{fig:method} as an example. We first determine node 0's interacting neighbor item sequence $\mathcal{V} = \{ v_1, \ldots, v_i, v_j, \ldots, v_n \}$, behavior sequence $\mathcal{B} = \{ b_1, \ldots, b_i, b_j, \ldots, b_n \}$, and timestamp $\mathcal{T} = \{ t_1, \ldots, t_i, t_j, \ldots, t_n \}$ before time $t$. The item sequence $\mathcal{V}$ and the behavior sequence $\mathcal{B}$ are used to generate the initial neighbor node representation $E^l(\mathcal{V})$ through Transformer-based multi-behavior correlation modeling. Then, combined with the time and behavior sequences as inputs, the neighbor node representation $Z(t) = \{ z_1, \ldots, z_i, z_j, \ldots, z_n \} $ and the initial target node representation $z_0(t)$ are generated through time-aware behavioral encoding. In this section, we will explain in detail how to aggregate the neighbor node sequence representation $Z(t)$ with the initial representation $z_0(t)$ of the target node to obtain the time-span and behavior-aware final representation ${h_0}^{(L+1)}(t)$ of the 1-hop target node.

Previously, we obtain the representation $z_0(t)$ of the target node and $Z(t)$ of the neighboring node sequence $\mathcal{V}$ through time-aware behavior encoding, which the neighbor node representation $Z(t)$ can be denoted as:
\begin{equation}\label{eq:Z(t)}
  \begin{aligned}
  Z(t) &= \left[ 
  \begin{array}{c}
    E^{l}(v_1) \| x_{0,1}(t_1) \| \Phi(0), \\
    \vdots \\
    E^{l}(v_n) \| x_{0,n}(t_n) \| \Phi(t-t_n), 
  \end{array}
  \right]
  \end{aligned}
\end{equation}
We then feed them into different linear projections for $q$, $k$, and $v$ decomposition:
\begin{equation}\label{eq:Query}
  \begin{aligned}
    q(t) &= [z_0(t)] W_Q, \\
    K(t) &= [Z(t)]_{1:n} W_K, \\
    V(t) &= [Z(t)]_{1:n} W_V
  \end{aligned}
\end{equation}
where $W_Q, W_K, W_V \in R^{(d_N+d_T+d_B) \times d}$ is the weight matrix to capture the interactions between node representations, behaviors, and time. Meanwhile, as shown in Eq(\ref{eq:K}), when calculating the raw attention score, the multiplication of $q(t)$ and $K(t)$ can effectively capture the time span information between nodes. Then, the neighborhood hidden representation $\tilde{h}_0(t) \in R^d$ of the target node obtained via self-attention mechanism is given by:
\begin{equation}\label{eq:h(t)}
  {\tilde{h}_0(t) = \text{Softmax}(\frac{q(t) {K(t)}^T}{\sqrt{d}})V(t)}
\end{equation}

Finally, we extend to the multi-head self-attention mechanism. Through residual connections\cite{residual} and layer normalization(LN)\cite{layernorm}, the final target node embedding can be denoted as:
\begin{equation}\label{eq:h00(t)}
  \begin{aligned}
    {h_0}^{(L+1)}(t) &=  \text{LN}\left( \text{FFN}\left( \tilde{h}_0^{(1)}(t) \| \ldots \| \tilde{h}_0^{(h)}(t) \| z_0 \right) + z_0 \right)
  \end{aligned}
\end{equation}
where $h$ represents the number of heads in the multi-head self-attention. In this way, we have completed a dynamic graph aggregation that incorporates both time span and behavioral awareness.

Moreover, by stacking time-aware aggregation layers, we can extend the graph up to $L$ hops. Considering that 1-hop graphs are insufficient for capturing higher-order associations between users or items, and 3-hop and above graphs are computationally expensive due to the exponential growth in the number of edges with each additional hop, we choose to use 2 hops.

\subsection{Boundary-Aware Personalized Prediction and Optimization}
In real-world scenarios, different user behaviors typically have distinct interest boundaries\cite{boundary}. To efficiently model these personalized behavioral boundaries during model optimization while considering dependencies between behaviors, 
we introduce an auxiliary score $b_{u,k}(t) \in R^1$ for each user's interaction behavior at the timestamp $t$. This score represents the user interest boundary under different behaviors. It can be either a parameter that is learned or computed from the embedding of users and edges with behavioral information, as shown in Eq(\ref{eq:boundary}):
\begin{equation}\label{eq:boundary}
  {b_{u,k}(t) = MLP(u(t) \| \sum_{j} {W^{\text{edge}}_k}(t) \cdot {W^{att}_{k,j}(t)}) }
\end{equation}
where $u(t) \in R^d$ is the embedding vector of user $u$ at the time $t$, $k$ denotes the behavior type of the $k$-th interaction, ${W^{\text{edge}}_k}(t) \in R^d$ is the embedding of the behavioral edges on the graph, $W^{att}_{k,j}(t) \in R^{d\times d}$ is the concatenated matrix of learned inter-behavioral dependencies in Eq(\ref{eq:attention weights}). Through the above design, we aim to capture the personalized boundaries of users' behavioral and temporal perceptions.

Our loss function takes a pointwise form, and the score between user $u$ and the positive item $p$ as well as the negative item $n$ can be denoted as:
\begin{equation}\label{eq:score}
  \begin{aligned}
    s(u,p) &= \text{FFN}({h_u}^{(L+1)}(t), {h_p}^{(L+1)}(t)), \\
    s(u,n) &= \text{FFN}({h_u}^{(L+1)}(t), {h_n}^{(L+1)}(t))
  \end{aligned}
\end{equation}

In conclusion, our loss function is formulated as follows:
\begin{equation}\label{eq:loss}
  \begin{split}
  \mathcal{L} = &- \sum_{(u,p) \in S_{u}} \ln\sigma(s(u,p) - b_{u,k}(t)) \\
  &- \alpha \sum_{(u,n) \notin S_{u}} \ln\sigma (1 - \sigma(s(u,n) - b_{u,k}(t)))
  \end{split}
\end{equation}
where $\sigma$ refers to the sigmoid operation and $\alpha$ is a hyperparameter to balance the positive item and negative item scores and the boundary distributions.

In this loss function, we aim to maximize the score of positive examples minus the boundary and minimize the score of negative examples minus the boundary. The loss function penalizes negative samples with scores above the boundary and positive samples with scores below the boundary. Meanwhile, $b_{u,k}$ models the interest boundaries for each user behavioral interaction, providing different boundaries for different user behaviors. By adjusting the hyperparameter $\alpha$ to balance the score distribution, we can find the optimal distribution. Algorithm~\ref{alg:Algorithm} explains the learning process of our MB-DGT framework.

\begin{algorithm}[htbp]
  \caption{The Proposed MB-DGT Algorithm}\label{alg:Algorithm}
  \KwIn{user set $\mathcal{U}$ = \{$u_i$\}, item set $\mathcal{V}$ = \{$v_j$\}, behavior set $\mathcal{B}$ = \{$b_k$\}, timestamp set $\mathcal{T}$ = \{$t_n$\}, interaction sequences $\mathcal{S}$, learning rate $\eta$, and number of epochs \textit{E}}
  \KwOut{trained model with parameters $\Theta$}
  Initialize parameters $\Theta$\;
  \For{$e$ = 1 to \textit{E}}{
      \For{$i$ = 1 to $|\mathcal{U}|$}{
          \For{$n$ = 1 to $|\mathcal{T}|$}{
              \For{$l$ = 1 to \textit{L}}{
                  Calculate the sequential representation based on the transformer-based multi-behavior correlation modeling according to (\ref{eq:qkv]})-(\ref{eq:f(i,j)})\;
                  Conduct the time-aware aggregation according to (\ref{eq:t})-(\ref{eq:h00(t)})\;
                  Conduct the boundary-aware personalized prediction according to (\ref{eq:boundary})-(\ref{eq:score})\;
              }
          }
      }
      Calculate loss $\mathcal{L}$ according to (\ref{eq:loss})\;
      \For{$\theta \in \Theta$}{
          $\theta = \theta - \eta \cdot \frac{\partial \mathcal{L}}{\partial \theta}$\;
      }
  }
\end{algorithm}

\subsection{Complexity Analysis}
\subsubsection{\textit{Time Complexity Analysis}} Due to the exponential growth in the number of edges brought about by the multi-hop graph structure, the time complexity of L-layer temporal aggregation can be expressed as \( O(n^L \cdot d) \). For transformer-based modeling of multi-behavior sequences, the time complexity of Eqs(\ref{eq:qkv]}) and (\ref{eq:final attention output}) is \( O(n^2 d + n d^2) \), where \( O(n^2 d) \) usually dominates. In our case, with \( L = 2 \), the final time complexity can be expressed as \( O(n^2 d) \), where the neighbor sequence length is $n$ and the model dimension is represented by $d$.

\subsubsection{\textit{Space Complexity Analysis}} For time-aware aggregation, each attention head requires only \( O((d_N + d_T + d_B) d_h + (d_h + d_0) d_f + d_f d) \) parameters, which is independent of neighborhood size. The transformer-based multi-behavioral sequence modeling requires \( O((|\mathcal{V}| + |\mathcal{E}|) d) \) for embedding, \( O(|\mathcal{B}|d^2)\) for attentional computation, and \( O(n) \) for relative behavioral encoding. The total overhead can be expressed as  \( O((d_N + d_T + d_B) d_h + (d_h + d_0) d_f + d_f d + (|\mathcal{V}| + |\mathcal{E}|) d + |\mathcal{B}|d^2 + n) \), where $d_h$ is the hidden layer dimension, $d_f$ is the FFN layer dimension, $d_0$ is the original embedding dimension of a node, and $d_N$, $d_T$ and $d_B$ represent the embedding dimension of node, time and behavior, respectively.
  
\section{Experiments}
In this section, we evaluate the effectiveness of the proposed MB-DGT model by performing extensive experiments using three real-world datasets. The experiments are designed to address the following research questions:

\begin{enumerate}[label=\textbullet\ RQ\arabic*: , leftmargin=3em, labelsep=0.2em]
    \item How does our model compare to the state-of-the-art methods in traditional recommendation, sequential recommendation, multi-behavior recommendation, and multi-behavior sequential recommendation?
    \item How does model's each designed component contribute to the overall performance?
    \item How do the various auxiliary behaviors affect the prediction performance on the target behavior?
    \item How to show that MB-DGT explicitly captures user-personalized behavioral boundaries?
    \item How is the interpretability of MB-DGT in capturing cross-type behavior dependencies?
    \item How does the performance of MB-DGT change with varying hyper-parameter values?
\end{enumerate}

\subsection{Experiments Settings}
\subsubsection{\textit{Datasets}} We assess our model using three datasets collected form the real-world.  Table~\ref{tab:Dataset} presents the behavior types and relevant statistics for these datasets.

\begin{table}[htpb]\centering
  \caption{Dataset Statistics}
  \label{tab:Dataset}
  \resizebox{\linewidth}{!}{%
  \begin{tabular}{l c c c c}
    \toprule
    Dataset & \#users & \#items & \#interactions & Behavior type \\
    \midrule
    Yelp & 19,800 & 22,734 & 1.4 $\times$ \(10^6\) & \{Tip, Dislike, Neutral, Like\} \\
    Taobao & 147,894 & 99,037 & 7.7 $\times$ \(10^6\) & \{Click, Favorite, Cart, Buy\} \\
    Tmall & 31,882 & 31,232 & 1.5 $\times$ \(10^6\) & \{Click, Favorite, Cart, Buy\} \\
    \bottomrule
  \end{tabular}}
\end{table}

\textit{Yelp:} This dataset is widely utilized for recommendation tasks and was collected from the Yelp challenge. User interactions are categorized based on explicit rating scores, which range from 1 (lowest) to 5 (highest) stars with increments of 0.5 stars. Specifically, interactions are classified into three behavior types: dislike (ratings in the range [0, 2]), neutral (ratings in the range (2, 4)), and like (ratings in the range [4, 5]). Additionally, the dataset contains timestamps for each interaction and allows users to provide tips about visited venues, which is considered a separate behavior. In this study, we focus on the target behavior ‘like’.

\textit{Taobao:} This dataset is derived from a real-world e-commerce platform and includes explicit user-item interactions across several types: click, add to cart, add to favorite, and buy. Each interaction also includes a timestamp, providing temporal context to the data. Among these interactions, the target behavior is treated as 'buy' because it's highly indicative of Gross Merchandise Volume (GMV) in online retail. GMV represents the overall revenue generated from merchandise sales and is a key metric for evaluating sales performance\cite{GMV1, GMV2}.

\textit{Tmall:} This dataset is sourced from one of the largest e-commerce platforms in China, Tmall. It encompasses various types of user behaviors, such as clicks, add items to favorites/carts, and buy, each associated with a timestamp. In this study, ‘buy’ is considered the target behavior. Consistent with the approach described in\cite{CML}, 
we only consider users who have made purchases a minimum of three times for inclusion in the training as well as the testing set.

\subsubsection{\textit{Evaluation Protocols}} We use two commonly adopted metrics, Hit Ratio (HR@N) and Normalized Discounted Cumulative Gain (NDCG@N), for ranked evaluations. Performance metrics are reported for N = 5 and 10 ranked cutoffs. For training test set partitioning, referring to  the methods\cite{MB-STR, PBAT}, we use a leave-one-out strategy. Specially, in the test set, the final interaction in each sequence, representing the target behavior, is used as a positive example, and the rest of interactions make up the training set. Following the evaluation approach used for above methods, we randomly selected 99 items from the entire item set as negative samples to be ranked and evaluated alongside the positive sample.


\subsubsection{\textit{Baselines}} To validate the effectiveness of the MB-DGT method, we conduct comparisons with several baseline approaches from different lines of research topics, covering i) traditional recommendation models (DMF, NGCF), ii) sequential recommendation models (GRU4Rec, Caser, SASRec, BERT4Rec), iii) multi-behavior recommendation models (NMTR, MB-GCN, MB-GMN), iv) multi-behavior sequential recommendation models (DIPN, BINN, MGNN-SPred, DMT, MBHT, MB-STR).

Traditional Recommendation Models: Traditional recommendation models use matrix factorization or collaborative filtering methods to explore the collaborative signals between users and items.
\begin{itemize}[]
  \item \textit{DMF}\cite{DMF} utilizes matrix decomposition to derive shared low-dimensional representations for user/item embedding.
  \item \textit{NGCF}\cite{NGCF} explores collaborative signals and high-order correlations in integration graphs.
\end{itemize}

Sequential Recommendation models: Sequential recommendations utilize users' historical sequence data
to capture the short and long term preferences, forecasting items the user is probably to engage with in the future.
\begin{itemize}[]
  \item \textit{GRU4Rec}\cite{GRU4Rec} leverages gate recurrent units to model temporal information in user sessions and encode sequential information.
  \item \textit{Caser}\cite{Caser} employs CNNs to capture the short and long term preferences and sequential signals.
  \item \textit{SASRec}\cite{SASRec} is a foundational model based on transformers that learns the pattern of sequential interaction for subsequent recommendations.
  \item \textit{BERT4Rec}\cite{Bert4rec} adopts a bi-directional transformer to make recommendations by fusing information from both sides of each item in user history.
\end{itemize}

Multi-Behavior Recommendation models: Multi-behavior recommendation models general embed users and items as vectors according to their previous multi-behavioral interactions to make recommendations.
\begin{itemize}[]
  \item \textit{NMTR}\cite{NMTR} proposes a multi-task framework to model the correlations among various behaviors in a hierarchical manner.
  \item \textit{MB-GCN}\cite{MB-GCN} employs GCNs to rebuild various user-item interactions by incorporating behavior-aware messaging during propagation.
  \item \textit{MB-GMN}\cite{MB-GMN} proposes a meta-learning paradigm to model multi-behavior pattern in a graph meta network.
\end{itemize}

Multi-Behavior Sequential Recommendation models: Multi-behavior sequential recommendation models simultaneously capture both users' heterogeneous behaviors and the sequential signals in behavior sequences, resulting in more accurate recommendations that align with users' actual preferences.
\begin{itemize}[]
  \item \textit{DIPN}\cite{DIPN} extracts user intents by leveraging multi-task learning and layered attention mechanism.
  \item \textit{BINN}\cite{BINN} maps items into a unified representation and selectively captures behavior information through LSTM architectures.
  \item \textit{MGNN-SPred}\cite{MGNN-SPerd} builds multi-relational item graphs with sub-sequences tailored to specific behaviors and combines representations using gating mechanisms.
  \item \textit{DMT}\cite{DMT} simultaneously models users' multiple behavior sequences with Transformers using MMoes to optimize objectives.
  \item \textit{MBHT}\cite{MBHT} introduces a multi-scale transformer augmented with hypergraphs to capture the short-term and long-term dependencies across different behavior types. 
  \item \textit{MB-STR}\cite{MB-STR} is a modern transformer-based model designed to capture behavior-specific signals and multi-behavior dependencies by modeling heterogeneous item-level correlations and using a novel positional encoding function.
\end{itemize}

\subsubsection{\textit{Parameter Settings}} We utilize Pytorch\cite{pytorch} to implement our proposed model. We use a Gaussian distribution \( \mathcal{N}(0, 0.02) \) to initialize parameters and optimize with Adam\cite{adam}. And the learning rate is set to 0.0001. 
To ensure fairness, we use the default optimal parameters provided by the authors of the baseline algorithms to replicate their reported results. We set the maximum training rounds to 50 and employ a multi-head mechanism with 2 heads. Furthermore, we vary the embedding dimension $d$  in \{60, 80, 100, 120, 140\}, and the neighboring nodes number $n$ in \{5, 10, 15, 20, 30\} to achieve the best performance. We ultimately set $d$ to 80 for Yelp and 100 for Taobao and Tmall. For the number of neighbors $n$, we set $n$=30 for Yelp, 15 for Tmall, and 10 for Taobao. To prevent the adverse effects of overfitting, early stopping strategy is used to stop training when Recall@10 on the test set fails to improve for three consecutive epochs.

\subsection{Performance Comparison (RQ1)} 
\begin{table*}[htbp]
  \centering
  \caption{Experimental results from three datasets are presented. The highest results are marked in bold, whereas the second-highest results are indicated with underlining.}\label{tab:performance}
  \renewcommand{\tabcolsep}{0.5pt} 
  \begin{tabularx}{\textwidth}{@{}l*{12}{>{\centering\arraybackslash}X}@{}}
  \toprule
  & \multicolumn{4}{c}{\textbf{Yelp}} & \multicolumn{4}{c}{\textbf{Taobao}} & \multicolumn{4}{c}{\textbf{Tmall}} \\
  \cmidrule(r){2-5} \cmidrule(lr){6-9} \cmidrule(l){10-13}
  & HR@5 & NDCG@5 & HR@10 & NDCG@10 & HR@5 & NDCG@5 & HR@10 & NDCG@10 & HR@5 & NDCG@5 & HR@10 & NDCG@10 \\
  \midrule
  DMF & 0.655 & 0.415 & 0.743 & 0.474 & 0.189 & 0.135 & 0.298 & 0.185 & 0.197 & 0.131 & 0.309 & 0.163 \\
  NGCF & 0.674 & 0.428 & 0.778 & 0.495 & 0.196 & 0.130 & 0.318 & 0.197 & 0.201 & 0.137 & 0.314 & 0.173 \\
  \midrule
  GRU4REC & 0.684 & 0.432 & 0.781 & 0.495 & 0.241 & 0.176 & 0.368 & 0.215 & 0.421 & 0.336 & 0.544 & 0.362 \\
  Caser & 0.673 & 0.425 & 0.763 & 0.489 & 0.204 & 0.132 & 0.341 & 0.193 & 0.393 & 0.293 & 0.513 & 0.325 \\
  SASRec & 0.728 & 0.451 & 0.793 & 0.501 & 0.253 & 0.182 & 0.377 & 0.226 & 0.432 & 0.337 & 0.557 & 0.381 \\
  BERT4Rec & 0.745 & 0.473 & 0.824 & 0.528 & 0.278 & 0.193 & 0.396 & 0.241 & 0.461 & 0.369 & 0.563 & 0.403 \\
  \midrule
  NMTR & 0.694 & 0.445 & 0.785 & 0.467 & 0.183 & 0.159 & 0.328 & 0.168 & 0.213 & 0.197 & 0.362 & 0.215 \\
  MB-GCN & 0.734 & 0.468 & 0.807 & 0.513 & 0.247 & 0.172 & 0.362 & 0.215 & 0.249 & 0.184 & 0.381 & 0.213 \\
  MB-GMN & 0.736 & 0.525 & 0.862 & 0.580 & 0.394 & 0.221 & 0.498 & 0.304 & 0.323 & 0.232 & 0.454 & 0.268 \\
  \midrule
  DIPN & 0.557 & 0.452 & 0.683 & 0.517 & 0.256 & 0.157 & 0.332 & 0.184 & 0.251 & 0.176 & 0.323 & 0.207 \\
  BINN & 0.375 & 0.344 & 0.489 & 0.413 & 0.358 & 0.224 & 0.463 & 0.306 & 0.393 & 0.237 & 0.463 & 0.289 \\
  MGNN-SPred & 0.372 & 0.335 & 0.485 & 0.402 & 0.353 & 0.218 & 0.459 & 0.304 & 0.377 & 0.231 & 0.438 & 0.275 \\
  DMT & 0.549 & 0.471 & 0.662 & 0.525 & 0.547 & 0.355 & 0.654 & 0.408 & 0.449 & 0.346 & 0.581 & 0.389 \\
  MBHT & 0.727 & 0.550 & 0.861 & 0.594 & 0.687 & \underline{0.590} & 0.764 & \underline{0.615} & 0.631 & 0.531 & 0.723 & 0.561 \\
  MB-STR & \underline{0.746} & \underline{0.569} & \underline{0.870} & \underline{0.610} & \underline{0.691} & 0.585 & \underline{0.768} & 0.608 & \underline{0.667} & \underline{0.547} & \underline{0.763} & \underline{0.577} \\
  \midrule
  \textbf{MB-DGT} & \textbf{0.765} & \textbf{0.606} & \textbf{0.879} & \textbf{0.645} & \textbf{0.705} & \textbf{0.605} & \textbf{0.781} & \textbf{0.629} & \textbf{0.688} & \textbf{0.569} & \textbf{0.774} & \textbf{0.596} \\
  Impv. & 2.55\% & 6.50\% & 1.04\% & 5.74\% & 2.03\% & 2.54\% & 1.69\% & 2.27\% & 3.15\% & 4.02\% & 1.44\% & 3.29\% \\
  \bottomrule
  \end{tabularx}
  \end{table*}

We assess the effectiveness of target behavior prediction by comparing all baseline models with our MB-DGT model. The results for three datasets are summarized in Table~\ref{tab:performance}. From this comparison, we can draw the following conclusions:
\begin{itemize}
    \item MB-DGT shows its effectiveness for multi-behavior sequential recommendation problem. The model delivers superior performance on every metric and across all datasets when compared to all baseline methods, which can be attributed to: i) MB-DGT captures multi-behavioral dependencies within sequences through transformer-based multi-behavioral correlation modeling, enabling better identification of item importance across different behaviors. ii) By constructing a 2-hop multi-behavioral user-item temporal interaction graph and applying time-aware aggregation, MB-DGT effectively captures dynamic changes in user interests within sequences and higher-order connectivity between sequences. iii) By learning the interest boundaries of user behaviors, MB-DGT can make more accurate recommendations for target behaviors.
    \item Both sequential and multi-behavior approaches improve model performance. Capturing sequential dependencies improves recommendation effectiveness. Generally, sequence recommendation baselines outperform most traditional baselines that do not utilize sequential information. The performance comparison between multi-behavioral sequence recommendation and multi-behavioral recommendation also demonstrates the effectiveness of incorporating sequential information. Additionally, leveraging multi-behavioral information enhances the modeling of user preferences. Multi-behavior recommendation usually outperforms traditional methods, and multi-behavior sequential recommendation often enhances prediction performance compared to sequential recommendation methods.
    \item MB-DGT consistently outperforms multi-behavior sequential recommendation based methods. Firstly, compared to models based on RNNs, such as DIPN and BINN, and those based on GNNs, like MGNN-SPred, MB-DGT fully leverages the transformer-based multi-behavior coorelation modeling to capture short and long term multi-behavioral sequential dependencies. Secondly, compared to models based on Transformers, such as DMT, MBHT, and MB-STR, MB-DGT conveys higher-order connectivity and effectively models dynamics through a multi-hop multi-behavioral interaction graph of user items. 
\end{itemize}

\subsection{Model Ablation Study (RQ2)}
\begin{table*}[htbp]
  \centering
  \caption{Performance analysis of MB-DGT via ablation study of key components.}\label{tab:ablation}
  \renewcommand{\tabcolsep}{0.5pt} 
  \begin{tabularx}{\textwidth}{@{}l*{12}{>{\centering\arraybackslash}X}@{}}
  \toprule
  & \multicolumn{4}{c}{\textbf{Yelp}} & \multicolumn{4}{c}{\textbf{Taobao}} & \multicolumn{4}{c}{\textbf{Tmall}} \\
  \cmidrule(r){2-5} \cmidrule(lr){6-9} \cmidrule(l){10-13}
  & HR@5 & NDCG@5 & HR@10 & NDCG@10 & HR@5 & NDCG@5 & HR@10 & NDCG@10 & HR@5 & NDCG@5 & HR@10 & NDCG@10 \\
  \midrule
  w/o TMCM and BAPP & 0.652 & 0.536 & 0.792 & 0.581 & 0.246 & 0.204 & 0.313 & 0.225 & 0.395 & 0.324 & 0.490 & 0.354 \\
  w/o TMCM & 0.692 & 0.572 & 0.814 & 0.612 & 0.264 & 0.232 & 0.362 & 0.264 & 0.431 & 0.369 & 0.515 & 0.397 \\
  \midrule
  w/o TAA and BAPP & 0.681 & 0.488 & 0.839 & 0.541 & 0.524 & 0.393 & 0.651 & 0.434 & 0.492 & 0.361 & 0.641 & 0.408 \\
  w/o TAA & 0.702 & 0.507 & 0.845 & 0.554 & 0.587 & 0.467 & 0.693 & 0.501 & 0.550 & 0.411 & 0.685 & 0.454 \\
  \midrule
  w/o BAPP & 0.729 & 0.578 & 0.856 & 0.619 & 0.653 & 0.536 & 0.748 & 0.566 & 0.627 & 0.504 & 0.742 & 0.542 \\
  \midrule
  \textbf{MB-DGT} & \textbf{0.765} & \textbf{0.606} & \textbf{0.879} & \textbf{0.645} & \textbf{0.705} & \textbf{0.605} & \textbf{0.781} & \textbf{0.629} & \textbf{0.688} & \textbf{0.569} & \textbf{0.774} & \textbf{0.596} \\
  \bottomrule
  \end{tabularx}
\end{table*}

To evaluate the effectiveness of the three main components, we conduct ablation experiments on various modified versions of the model: 

\begin{itemize}
  \item w/o TMCM and BAPP: We remove the TMCM and BAPP modules, using the initial representations of nodes for time-aware aggregation without capturing sequential behavioral dependencies or modeling user behavioral interest boundaries.
  \item w/o TMCM: We remove the TMCM module and use the initial node representations for time-aware aggregation without capturing sequential behavioral dependencies, but still model user behavioral interest boundaries during optimization.
  \item w/o TAA and BAPP: We remove the TAA and BAPP modules. Instead of performing time-aware dynamic graph aggregation, we directly use the captured multi-behavioral sequence information as node representations after applying the MLP, disregarding user behavioral interest boundaries.
  \item w/o TAA: We remove the TAA module and model user behavioral interest boundaries without considering time-aware dynamic graph aggregation, using the multi-behavioral sequence information characterized by the MLP as node representations.
  \item w/o BAPP: We remove the BAPP module, allowing the neighbor nodes to go through the TMCM module to capture sequential behavioral dependencies before completing time-aware aggregation through the TAA module. However, user behavioral interest boundaries are not modeled during optimization.
\end{itemize}

Through the results in Table~\ref{tab:ablation},  we can observe that: i) MB-DGT outperforms in three variants: w/o TMCM, w/o TAA and w/o BAPP, indicating the effectiveness of three main components. Comparing w/o TMCM and w/o TAA, w/o TMCM has a worse performance on two datasets, suggesting that in most cases, the  pattern of behavior changes plays a more critical role than timing intervals in capturing the complex collaborations inherent in multi-behavior sequences. ii) Pairwise comparisons between w/o TMCM and BAPP and w/o TMCM, w/o TAA and BAPP and w/o TAA, and w/o BAPP and MB-DGT demonstrate the applicability and effectiveness of our proposed user behavioral interest boundary modeling optimization. This method establishes personalized decision boundaries for various user behaviors, enhancing the accuracy of model predictions.


\subsection{Effect of Auxiliary Behaviors (RQ3)}
\begin{figure}
  \centering
  \includegraphics[width=1\linewidth]{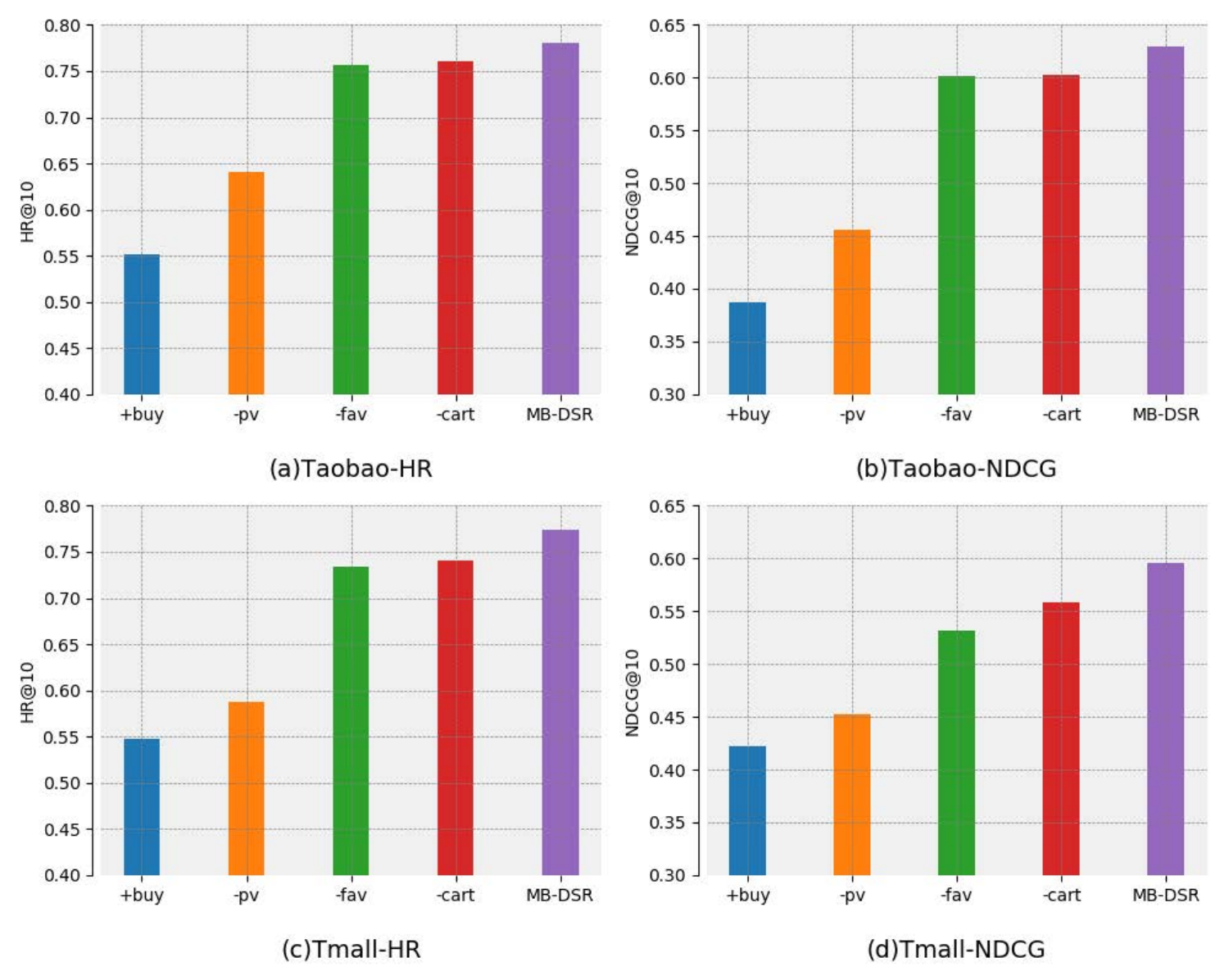}\\
  \caption{The performence of auxiliary behavioral data ablation experiments on HR@10 and NDCG@10.}\label{fig:fig4}
\end{figure}
We also perform an auxiliary behavioral data ablation study to evaluate the impact of various auxiliary behaviors on model performance. We design four filters for the training data and then examine the ablation effect of different behaviors on two datasets Taobao and Tmall: '-click', '-fav', and '-cart' indicate that we filter out interaction records in the training set whose behavior is click, like and add to cart, respectively. Additionally, 'only buy' indicates that we only use records of buy behavior for recommendations. Fig~\ref{fig:fig4} shows the evaluation results.

We can conclude from the result that: i) MB-DGT outperforms all variants, demonstrating 
the efficacy of utilizing multiple behavioral data in modeling user preferences. ii) Performance worsens when any form of auxiliary behavior is removed, with the most significant drop occurring when page view is excluded. 
This suggests that page views serve as a crucial factor in generalizing behavior patterns and exhibit a strong correlation with the target behavior. 
This also aligns with the observation that page view behavior generally happens more frequently.

\subsection{User Behavior Interest Boundaries (RQ4)}

\begin{figure}
  \centering
  \includegraphics[width=1\linewidth]{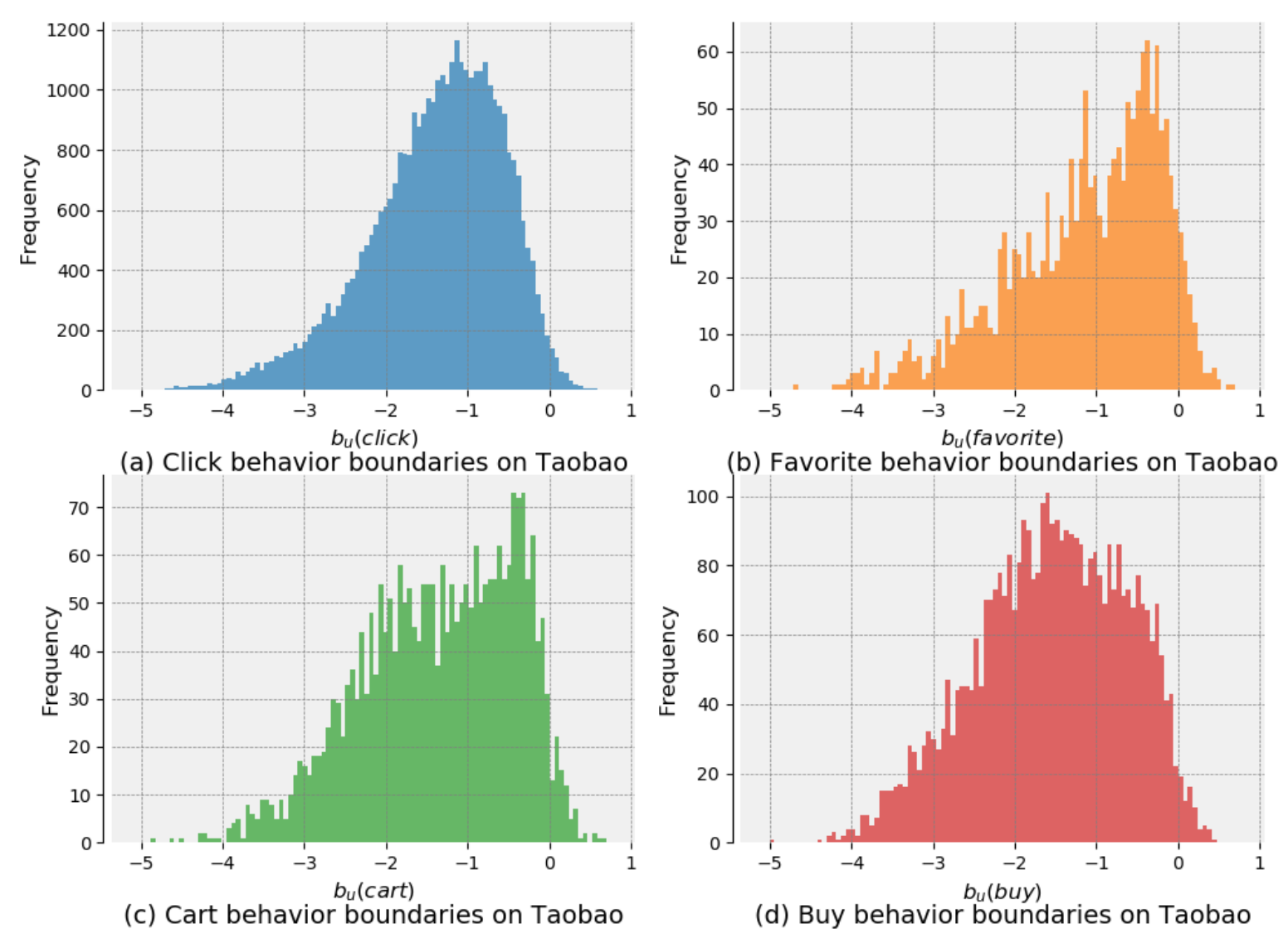}\\
  \caption{Boundary distribution on different behaviors for one thousand users on Taobao dataset.}\label{fig:fig5}
\end{figure}

To assess if the model assigns different boundaries for various user behaviors, Fig~\ref{fig:fig5} presents the distribution of learned boundaries $b_{u,k}(t)$ for four behaviors across one thousand users in the Taobao dataset. The results confirm that our model assigns distinct boundaries for different user behaviors, represented as different normal distributions.

\begin{figure}
  \centering
  \includegraphics[width=1\linewidth]{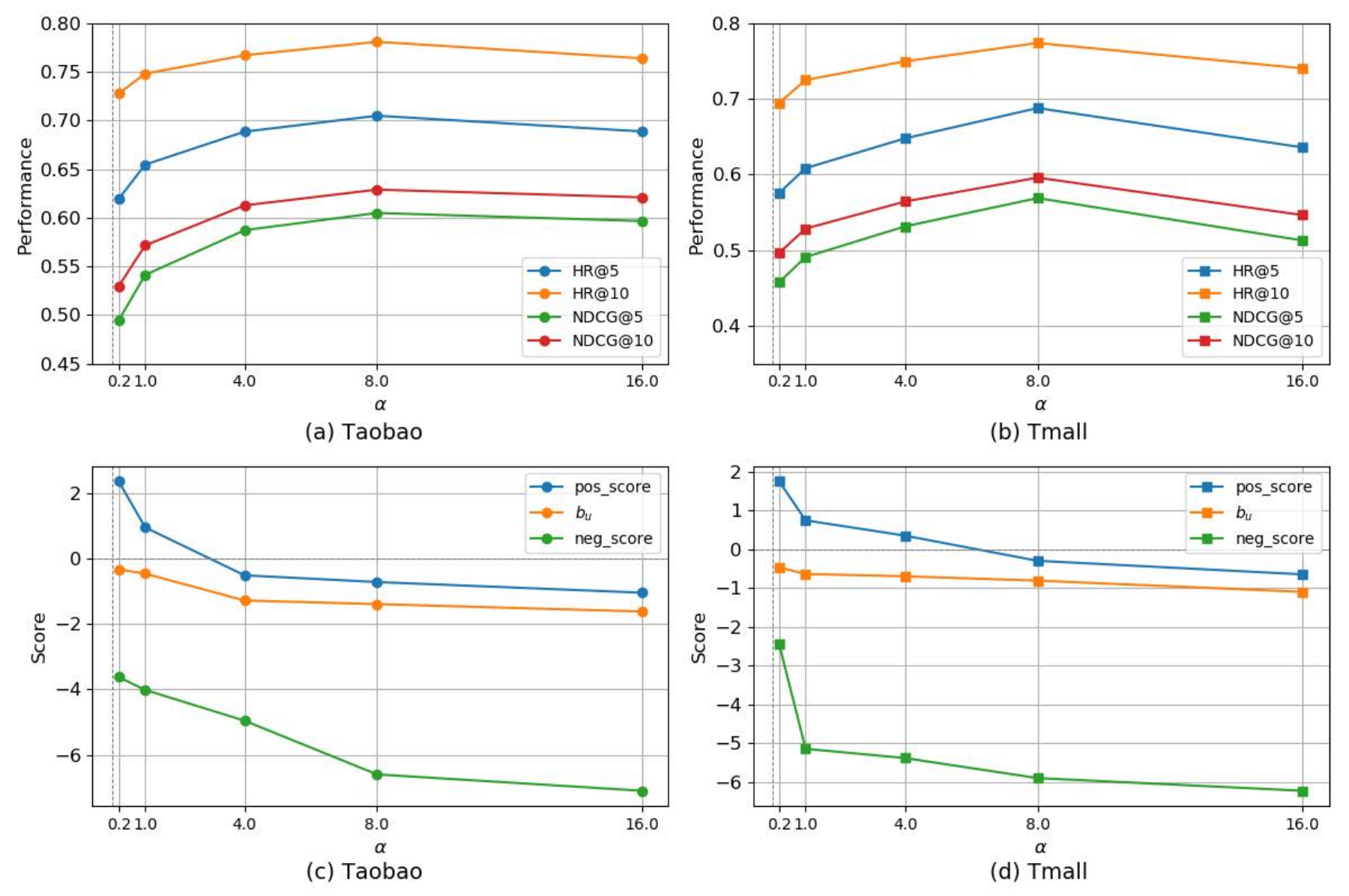}\\
  \caption{The model performance with different settings of $\alpha$, as well as the mean values of $s(u,p)$, $s(u,n)$, and $b_{u,k}(t)$ under various $\alpha$ settings}\label{fig:fig6}
\end{figure}

Moreover, our approach introduces only one hyperparameter, $\alpha$, to adjust the positive and negative samples scores distribution. To investigate the effect of $\alpha$ in our model, we conduct experiments varying $\alpha \in \{0.2, 1, 4, 8, 16\}$ settings on the Taobao and Tmall datasets. In addition to analyzing the variation of the learned boundaries, we also examine the changes in positive and negative case scores for different values of $\alpha$. As Fig. 6 illustrates, we can deduce that: i) From the top two plots of Fig~\ref{fig:fig6}, it is confirmed that the model performance is affected by $\alpha$ settings , with an optimal $\alpha$ achieving the best results. ii) As $\alpha$ increases, the boundary $b_u$ consistently decreases, indicating a strong correlation between $\alpha$ and boundary distribution. This implies that increasing $\alpha$ to give more weight to the negative loss component effectively reduces the boundary, influencing both positive and negative aspects concurrently. iii) As $\alpha$ increases, negative sample scores are increasingly situated further from the boundary. This occurs because a larger $\alpha$ makes the negative sample loss a more significant factor, causing $\sigma(s(u,n)-b_{u,k})$ to approach zero.


\subsection{Model Interpretation Study (RQ5)}
\begin{figure}
  \centering
  \includegraphics[width=1\linewidth]{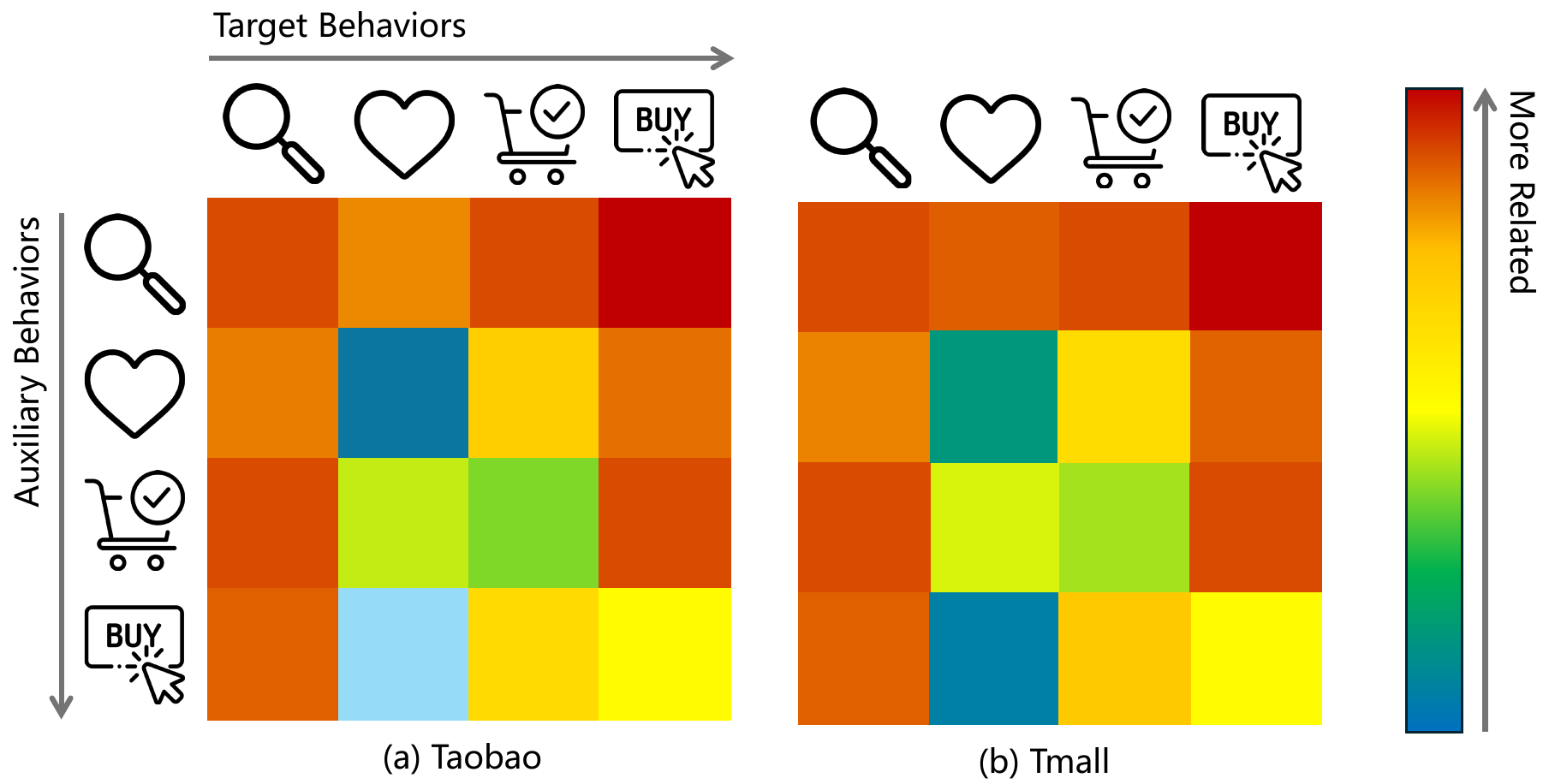}\\
  \caption{Visualization of multi-behavioral dependencies learned by MB-DGT. For each behavior on the vertical axis, we assess the relevance of all other behaviors in relation to it. The intensity of the square colors indicates the degree of these relevances.}\label{fig:fig7}
\end{figure}

In our framework, all types of behaviors can be classified as either auxiliary or target behaviors. Specially, Fig~\ref{fig:fig7} shows the behavioral patterns captured by MB-DGT on the Taobao and Tmall datasets. For every behavior on the vertical axis, we analyze the correlations with other behaviors, represented by the color shades of the squares. We generated a 4×4 dependency matrix corresponding to the average attn scores of all test users and their behaviors on the test set. The visualization results reveal hierarchical and interpretable correlations between different behavioral types (4 types). For example, in the Taobao dataset, the ‘buy’ behavior shows a high correlation with ‘page view’ and ‘add to cart,’ but a low correlation with ‘favorites.’ Similar conclusions can be drawn from the Tmall dataset results. It is confirmed that our model can identify the importance of items across various behaviors.

\subsection{Parameter analysis (RQ6)}
Now, we examine how hyper-parameters affect the performance of our model, specifically focusing on the embedding dimension $d$, and the number of neighbors $n$. 

\begin{figure}
  \centering
  \includegraphics[width=1\linewidth]{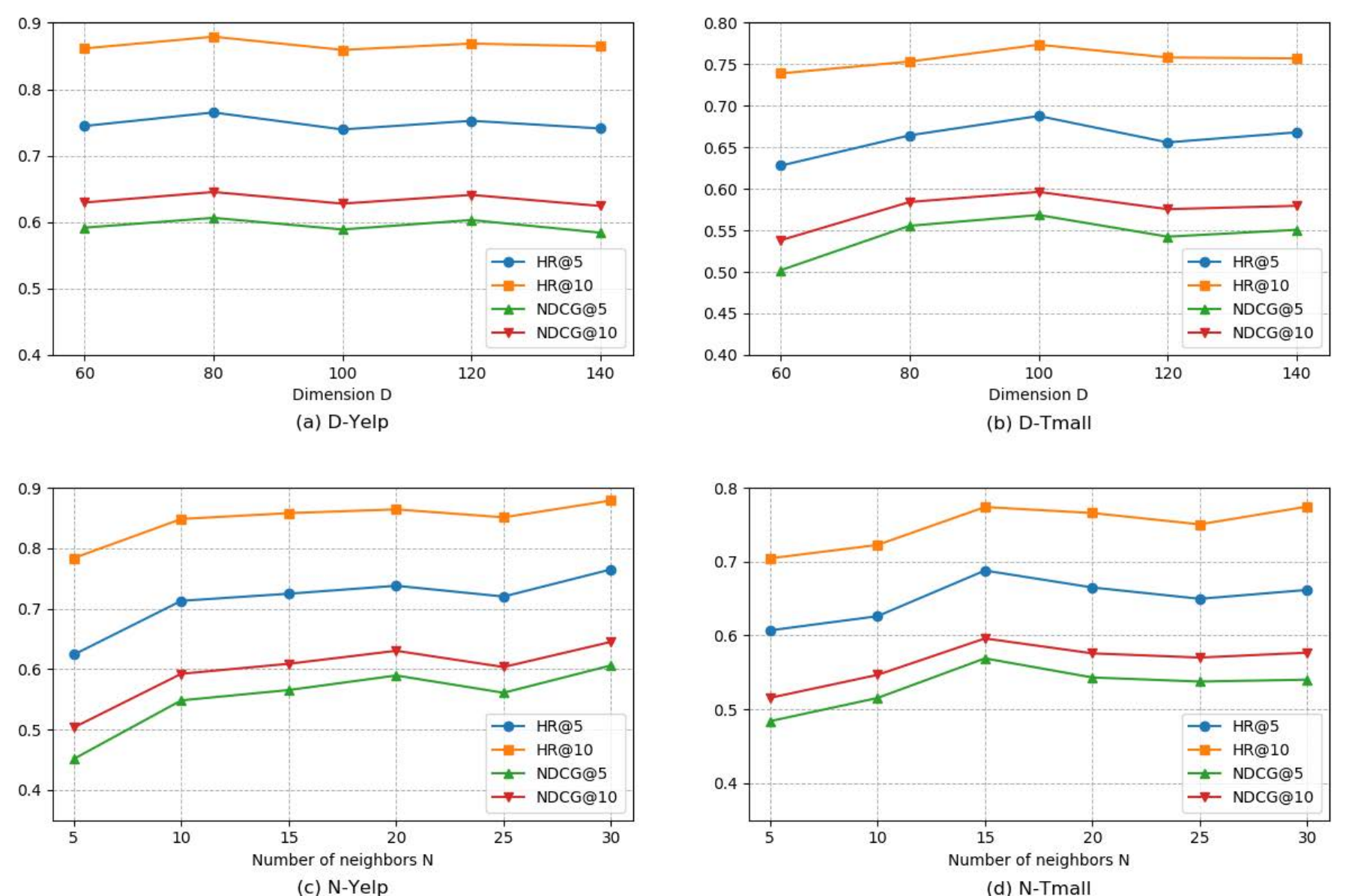}\\
  \caption{Hyperparameter study on MB-DGT, including model dimensionality $d$ and number of neighbors $n$, evaluated on the Yelp and Taobao datasets.}\label{fig:fig8}
\end{figure}

We first examine the effect of model dimension on MB-DGT by varying $d$ in $\{60, 80, 100, 120, 140\}$ and reporting performances on Yelp and Tmall in Fig~\ref{fig:fig8}(a)-(b). The results indicate that an appropriate increase in the model dimension can improve the model's effectiveness, but too high a dimension may contain too much redundant information triggering a decrease in effectiveness and an increase in computational overhead. Considering the effects and the training computational cost, we finally choose $d$ = 80 for Yelp, $d$ = 100 for Taobao and Tmall, respectively.

Subsequently, we conduct experiments by altering the number of neighbors $n$ in $\{5, 10, 15, 20, 25, 30\}$. Fig~\ref{fig:fig8}(c)-(d) show the model performance under various $n$ on two datasets, indicating that increasing the number of neighbors on the Yelp dataset generally achieves an increase in effectiveness, but too large a number of neighbors on the Tmall dataset may lead to a decrease in effectiveness. The reason may be that Yelp has longer average user interaction sequences and more dispersed behaviors, so a larger number of neighbors is needed to aggregate information, while Tmall has shorter user interaction sequences and more concentrated behaviors, so a smaller number of neighbors can aggregate useful information more accurately. In this case, we choose $n$ = 30 for Yelp, $n$ = 15 for Tmall and $n$ = 10 for Taobao to effective model training. In addition, our model effectively addresses the cold-start problem. Compared to other multi-behavior sequential recommendation models\cite{MB-STR,PBAT}, we require significantly fewer neighbor sequences. For instance, while other models typically need 50 neighbor sequences, our model achieves or even exceeds their performance with as few as 10 neighbor sequences. This efficiency is due to our ability to capture higher-order associations between sequences.

\section{Conclusion}
In this study, we introduce boundary-aware Multi-Behavior Dynamic Graph Transformer (MB-DGT) model, a novel approach to improve sequential recommendation under multi-behavior scenarios. Specifically, we develop a transformer-based multi-behavior correlation model to extract heterogeneous user interests for each behavior type. Furthermore, evolving user interests are comprehensively modeled at different points in time through time-aware aggregation. For model optimization, we model explicit scores for different user behavior interest boundaries and enhance recommendation accuracy. Comprehensive experiments carried out on three publicly available datasets confirm the effectiveness of MB-DGT. In the future, we plan to enhance the modeling efficiency of MB-DGT to more effectively capture dynamic user interests using multi-behavioral information. Additionally, we will further explore the model's recommendation capabilities under sparse data and cold-start conditions, aiming to improve its generalization performance and interpretability across different datasets.

\section*{Acknowledgments}
This work was supported by the National Natural Science Foundation of China No. 62102421 and  No. 62102035,  and Fundamental Research Funds for the Central Universities No. 2233100004.



\bibliographystyle{IEEEtran}
\bibliography{IEEEabrv,IEEEexample}

\vspace{11pt}


\vfill

\end{document}